\documentclass[aps,prb,twocolumn,floats]{revtex4-2}
\usepackage{epsfig}
\usepackage{amsmath}
\usepackage{amssymb}
\usepackage{graphicx}
\usepackage{soul}
\usepackage{color}

\begin{document}
\title{Floquet Analysis on an Irradiated Nodal Surface Semimetal with Non-Symmorphic Symmetry}
\author{Bhaskar Pandit$^1$, Satyaki Kar$^{2}$}
\email{Corresponding Author: satyaki.phys@gmail.com}
\affiliation{$^1$Netaji Mahavidyalaya, Arambagh, West Bengal - 712601, India\\$^2$AKPC Mahavidyalaya, Bengai, West Bengal -712611, India}
\begin{abstract}
  A nodal surface semimetal (NSSM) features symmetry enforced band crossings along a surface within the three-dimensional (3D) Brillouin zone (BZ) and a presence of a nonsymmorphic symmetry there pushes such surfaces to stick to the BZ {center or} boundaries. The topological robustness of the same does not always come with nonzero Berry fluxes. We consider two such nodal surfaces (NS), one with zero and another with nonzero topological charges and investigate the effect of light irradiation on them. We find that depending on the state of polarization, one can obtain additional Weyl points/ nodal surfaces in the corresponding Floquet Hamiltonians. Particularly, using a simple two band spinless/spin polarized models with no spin orbit coupling, we emphasize the low energy behavior of the continuum Hamiltonians close to the band crossings and its evolution in a Floquet system in the high frequency limit. {In the Floquet system,} we also find the nodal surfaces to perish or {new multi Weyl points to get popped up} for different polarization scenario or different NSSM Hamiltonians. Our findings open up important avenues on what out of equilibrium NSSM systems can offer in many active fields including quantum computations. 
\end{abstract}
\maketitle                              
\section{Introduction}

Band crossing along a two dimensional (2D) surface with linear dispersion away from it is what it takes to form a nodal surface\cite{wu} or Weyl surface\cite{zhong} or Weyl nodal surface\cite{turker} which generally indicates topological stability in presence of small symmetry preserving perturbations of the non-interacting Bloch Hamiltonians\cite{turker,1807}. A 2D nodal surface can be a closed surface forming a Dirac nodal sphere or pseudo Dirac nodal sphere\cite{wang}. But it can also be distinct from conventional Fermi surfaces of ordinary metals with the nodal surface representing a torus (and hence non-deformable continually to a sphere)\cite{wu} with opposite edges connected via periodic boundaries of the Brillouin zone (BZ).

To name a few, existence of Nodal surfaces (NS) has been probed in systems like quasi-one dimensional crystals\cite{liang}, graphene networks\cite{zhong} or multiband superconductors with broken time-reversal symmetry\cite{timm}. Acoustic/photonic systems have also be designed where additional linear Weyl points {(WP)} result in a topologically charged NS\cite{yang,xiao}.
A nodal surface semimetal (NSSM) can be obtained as compulsion from the global internal symmetry ($eg.,$ one component of spin operator) providing topological protection because the crossed bands with different eigenvalues for symmetry operator do not hybridize to open a gap\cite{turker}.
One can also consider protection due to a combination of unitary and antiunitary symmetries, namely a space-time inversion symmetry $\mathcal {PT}$. This in addition to a sublattice symmetry $\mathcal S$ can carry a $\mathbb{Z}_2$ topological index\cite{wu} (identifying whether the band gap is inverted or not\cite{zhong}) for the NS. Independent $\mathcal{PT}$ and $\mathcal S$ operators with $[\mathcal {PT,S}]=0$, leads one to consider $\mathcal{PT=K}$, the complex conjugation, for the Bloch Hamiltonian $H(k)$ to be real whereas sublattice symmetry allows the Hamiltonian to be transformed to a block anti-diagonal form, which then can be unitary transformed to a real matrix\cite{wu}. Generically, band-crossing points can spread out a compact surface protected by the symmetries and one can find gapped points $k_{1,2}$ above and below that surface with opposite signs of the Pfaffian of the anti-diagonal Hamiltonian matrix (indicating topological charges) can be realized\cite{wu,turker}.
But such stringent requirement also needs regions within BZ to have inverted band ordering with different $\mathbb{Z}_2$ class\cite{wu}. Easier is to find the essential nodal surfaces in a system obeying a nonsymmorphic symmetry where linear band crossings occur at sub-BZ boundaries\cite{wu}.

A non-symmorphic symmetry\cite{furusaki} is a combination of a point group symmetry and half-lattice translation (but not as a product of a lattice translation and a point group operation\cite{liang}) that restrict the form of the band structure both locally and globally leading to unavoidable band crossings in the bulk\cite{zhao}. A material with such symmetry generally possess a two fold screw rotation and glide reflection symmetry given symbolically as $S_z=\{C_{2z}|{\bf T}_z=c/2\}$. A combination of ${\rm C}=\mathcal{PT}$ and $D=\mathcal{P}S_z$ symmetry can give rise to nodal surfaces. The compound symmetry $D$ turns $k_z\rightarrow-k_z$ and is respected only where $k_z=-k_z,~i.e.,$ at $k_z=0$ and $k_z=\pi$ where $D$ anticommutes with $\mathcal{PT}$\cite{liang}. This protects the twofold degeneracy there. If we instead consider the compound symmetry $D'=\mathcal{T}S_z$ then also, for a spinless case, one gets Kramer's two-fold degeneracy at each point of the $k_z=\pi$ plane and thus the energy bands stick together at the boundary of the BZ\cite{liang,wu}. These band crossings are topologically robust both locally and globally\cite{zhao}.
Thus the combination of two-fold screw rotation symmetry and time reversal symmetry features essential band crossings solely determined by symmetry. The 3D system can be considered as a collection of 1D chains parameterized by transverse quantum numbers $(k_x,k_y)$\cite{wu}. Each 1D subsystems has a twisted band structure (like in a Mobius strip) with single band crossing at the boundaries. 
Here nodal surfaces can be protected even in presence of SOC provided space-time inversion symmetry is violated\cite{wu}.

A Nielsen-Ninomiya no-go theorem\cite{nogo} predicts overall zero topological charge within a BZ. But a nodal surface (NS) at the BZ sub-boundaries does not get any symmetrically positioned partner with opposite topological charge to make the overall charge zero and so even being topologically protected (due to having twisted band crossings like in a Mobius strip) by a twofold screw axis and time-reversal symmetry\cite{wu,zhao}, these NSs possess zero $\mathbb{Z}$ charge of Berry fluxes\cite{xiao}. A non-symmorphic symmetry features a global topology with $\mathbb{Z}_2$ classification where the band crossing can be avoided only using doubled Hamiltonians in four-band theories\cite{wu,zhao}. But in a two-band theory, one can also have NS not at such high symmetry points and get characterized by a $\mathbb{Z}_2$ topological index in presence of $\mathcal{PT}$ and $\mathcal{S}$ symmetry\cite{wu}. Besides, it is possible to attach 1D or 2D topological charges leading to nontrivial surface states in these systems. M. Xiao $et~al.$\cite{xiao} showed that even in presence of additional Weyl points within the 3D BZ, one can have topologically charged nodal surfaces at BZ boundaries ($i.e.,$ at $k_z=\pi$) yet obeying the no-go theorem.

In this report we study both examples of chargeless and topologically charged NSSM under light irradiation and investigate their dynamic behavior using a Floquet-Magnus analysis\cite{eckardt}. After mentioning the Hamiltonian formulation in Section II and topological aspects in Section III, we discuss on the Floquet theory in the irradiated system in Section III. Lastly in Section  IV, we summarize our results and brief on further scopes of our work.

\section{Formulation}

Let us first consider an one dimensional chain. A two-fold unitary nonsymmorphic symmetry operator\cite{zhao} for a 2 band model there can be given as
\begin{displaymath}
  G(k_z)=\left(\begin{array}{cc}
    0 & e^{-ik_z}\\
    1 & 0
  \end{array}\right)
\end{displaymath}
with $G^2(k_z)=e^{-ik_z}\sigma_0$ and thereby having $G(k_z)$ eigenvalues of $\pm e^{-ik_z/2}$. The relation $GHG^{-1}=H$ requires the Hamiltonian to be of form
\begin{displaymath}
  H(k_z)=\left(\begin{array}{cc}
    0 & Q(k_z)   \\
    Q^\star(k_z) & 0
  \end{array}\right)
\end{displaymath}
with constraint $Q(k_z)e^{ik_z}=Q^\star(k_z)$. Now this requires $Q(k_z)$ to become zero for some momentum $k_z$\cite{zhao} indicating gaplessness in the system. In addition, presence of an inversion symmetry expressed via operator $\hat P=\sigma_2\hat i$ ($\hat P=\sigma_1\hat i$), where $\hat i$ inverses the momentum, leads to the symmetry enforced band crossings at $k_z=0$ (at BZ boundaries $i.e.,~k_z=\pm\pi$)\cite{zhao}, where the protected crossed energy bands have different $\hat P$ eigenvalues. In either case, one gets a twisted band structure like in a Mobius strip with band crossings at $k_z=0$ and $\pm\pi$, though only $\hat P=\sigma_1\hat i$, features crossed bands at the BZ sub-boundaries protected by nonsymmorphic symmetry. Hence this is an essential band crossing\cite{wu}.

Next for a magnetic nonsymmorphic symmetry (MNS), which is a combination of $G(k_z)$ and time reversal symmetry $T$, we find that for $T=\mathcal{K}\hat i$, two-fold band degeneracy among the Kramer's pairs occurs at every point with $k_z=\pi$.
One can see this from the commutation relation $[GT,H]=0$ leading to $e^{ik_z}Q(k_z)=Q(-k_z)$. This makes $Q(\pi)=0$ and thus $k_z=\pi$ a symmetry enforced band crossing point. Here one can also add a term $f(k_z)\sigma_z$ to the Hamiltonian with $f^\star(-k_z)=-f(k_z)$ satisfying the $GT$ symmetry. This demands $Re[f(\pi)]=0$.
Interestingly, the presence of nodal surfaces does not lead to any special gapless boundary mode for they have zero spatial codimensionality\cite{wu}. Typically a nodal surface is a torus in a 3D BZ.

Without loss of generality, we can consider a low energy continuum model of a NSSM to be defined about a point ${\bf k_0}=(k_{x0},k_{y0},\pi)$ within the flat nodal surface given by $k_z=\pi$. With $k\rightarrow k_0+q$ and
only considering the 1D subsystem along $\hat z$, one can write $H_{1D}(q_z)=H(\pi+q_z)$ for which MNS requires $-e^{iq_z}Q(\pi+q_z)=Q(\pi-q_z)$ and $f^\star(\pi+q_z)=-f(\pi-q_z)$.
  Here a Taylor's expansion up to 2nd order gives $Q''(\pi)=-iQ'(\pi)$ (and also $f(\pi+q_z)\sim\sin q_z$) for which a possible solution is $Q(\pi+q_z)\sim1-e^{-iq_z}$. This gives a low energy Hamiltonian $H_{1D}=v(1-\cos q_z)~\sigma_x-v\sin q_z~\sigma_y+v_z\sin q_z\sigma_z$ which for small $q_z$ becomes $H_{1D}=v\frac{q_z^2}{2}\sigma_x-vq_z\sigma_y+v_zq_z\sigma_z$.

Now for a 3D  system with nodal surfaces protected by nonsymmorphic symmetry, the 3D Hamiltonian
can be considered as a collection of 1D $k_z$ subsystems parameterized by $(k_x,k_y)$ so that one can write $H(k_x,k_y,k_z)=H_{1D}^{(k_x,k_y)}(k_z)$. A 3D Dirac Hamiltonian is expressed in terms of $4\times4$ matrices, though for our spinless system, a $2\times2$ Hamiltonian matrix suffices to describe the physics for merely an outer product of $\mathbb{I}_{2\times2}$ comes into play due to the spin subspace (or spin degeneracy).
Thus we can, in general, write the Hamiltonian as
\begin{eqnarray}
H_{NS}^{(1)}={c_1}(k_x,k_y)[\frac{q_z^2}{2}\sigma_x-q_z\sigma_y]+{d_1}(k_x,k_y)q_z\sigma_z.
\label{eq1}
\end{eqnarray}
For simplicity, one can consider constant prefactors {$c_1=d_1=1$} independent of $k_x$ and $k_y$ yielding a Hamiltonian in terms of variable $q_z$ alone.

\begin{figure}
  \begin{center}
   \begin{picture}(100,100)
     \put(-70,0){
    \includegraphics[width=.5\linewidth]{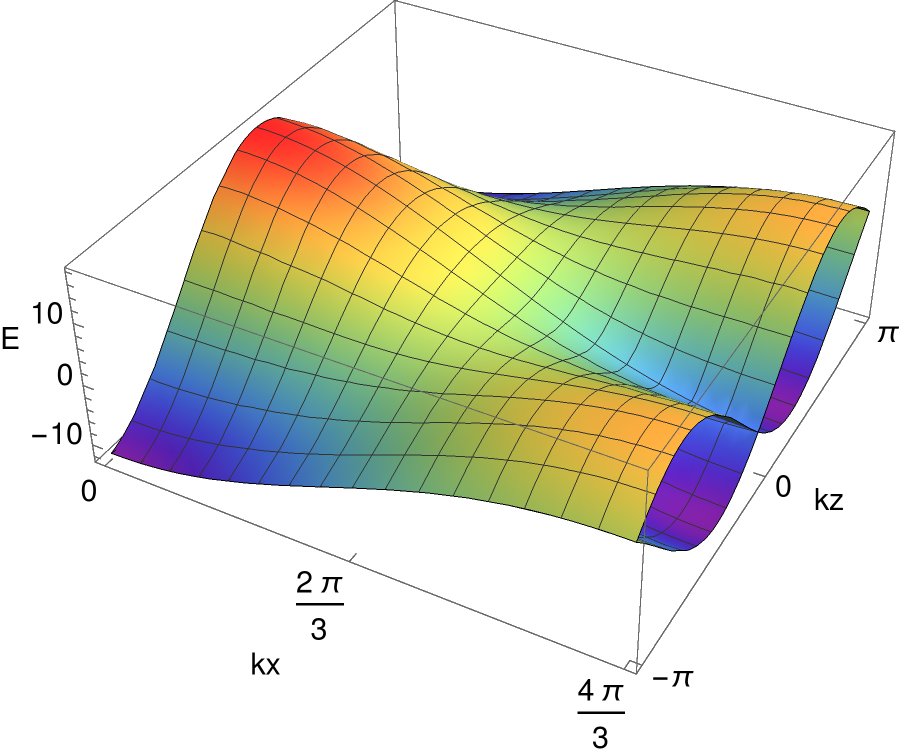}
    \includegraphics[width=.45\linewidth,height=1.4 in]{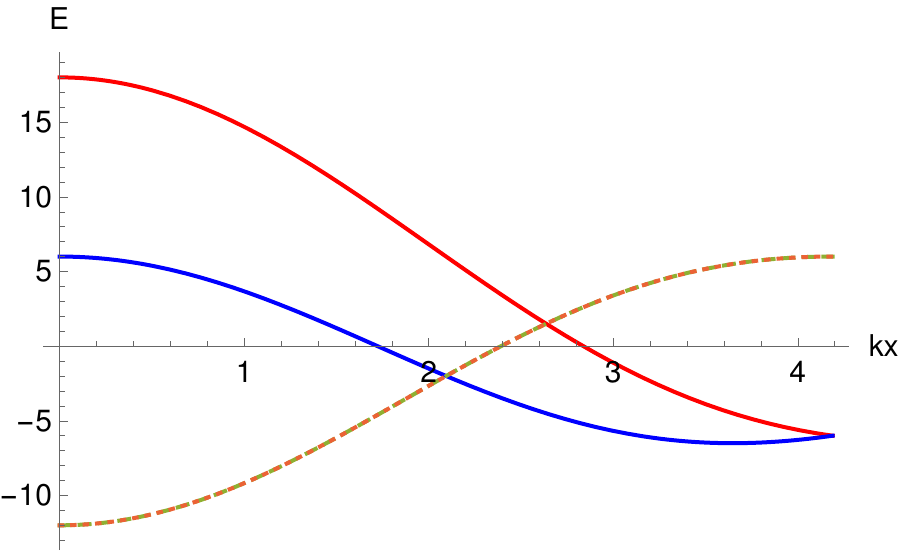}
    \put(-230,80){(a)}
    \put(-60,80){(b)}}
   \end{picture}
   \end{center}
   \vskip -0.2 in
\caption{{(a) Dispersions from the tight binding model (\ref{eqtb}) for $k_y=0$ of which (b) shows the cut along $k_x$ for $k_z=0$ (solid) and $k_z=\pi$ (dash-dotted). NS and WP can be identified from the gaplessness. Here we consider $t_0=t_c=1/2$.}} 
   \label{disp}
\end{figure}

This NSSM not necessarily feature nonzero topological charge.
However, one can choose smart tight binding models which feature both NS and Weyl points\cite{xiao} and in order to compensate the Weyl point charges, the NS requires to have nonzero topological charges (as per the no-go theorem). Xiao $et.~al.$ designed a phononic crystal featuring a time reversal and nonsymmorphic symmetric but inversion broken topologically charged NSSM\cite{xiao}. Likewise one can write a continuum Hamiltonian about a point $k_0=(k_{x0}, k_{y0}, \pi)$ in the nodal plane $k_z=\pi$ (with $k\rightarrow k_0 + q$) as\cite{xiao}
\begin{eqnarray}
H_{NS}^{(2)}={c_2}q_z(q_x\sigma_x+q_y\sigma_y) +{d_2}q_z\sigma_z
\label{eq2}
\end{eqnarray}
Thus the nodal surface is given by the $q_x-q_y$ plane for $q_z=0$. The prefactors {$c_2$ and $d_2$} are compatible with dimensionality and can be taken as unity for simplicity. To accommodate for nonzero charges, the 3D Brillouin zone (BZ) need to also incorporate {other topological entities ($e.g.,$ Weyl points)} with compensating topological charges as per the Nielsen-Ninomiya no-go theorem\cite{nogo}. In this regard, one can take a look at Ref.\cite{xiao} where the BZ has a pair of Weyl points at $(k_x=\pm4\pi/3,~k_y=0,~{k_z=0})$ apart from a NS at $k_z=\pi$ and they sprout from the tight-binding Hamiltonian
\begin{equation}\label{eqtb}
  H=\left(\begin{array}{cc}  h_1(k_z) & h_2\\ h_2^\star & h_1(-k_z) \end{array}\right)
\end{equation}
with $h_1(k_z)=2t_c[\cos(k_x+k_z)+2\cos(\frac{k_x}{2}-k_z)\cos(\frac{\sqrt{3}k_y}{2})]$ and $h_2=2t_0\cos(\frac{k_z}{2})[2\cos(\frac{k_x}{2})\exp(i\frac{\sqrt{3}k_y}{6})+\exp(-i\frac{\sqrt{3}k_y}{3})]$, with $t_c,~t_0$ being two hopping parameters.
{Notice that $h_2=0$ for $(k_x=\pm\frac{4\pi}{3},~k_y=0)$ and for $k_z=\pi$, the corresponding dispersions being given as $E=h_1(k_z),~h_1(-k_z)$. Thus band crossing occurs at points $(k_x=\pm\frac{4\pi}{3},~k_y=0,~k_z=0)$ and at the plane $k_z=\pi$. However, $E(k_z=\pi)$ at the NS varies with $k_x$ and $k_y$ and can be greater, smaller or same as energy $E(\pm\frac{4\pi}{3},0,0)$ at the WP (see Fig.\ref{disp}).} {Also notice that for $k_x=\pi/2$, $E=h_1(k_z)\pm|h_2|$ with no band crossing in general.}

In the following sections we discuss on the calculation of the topological invariants namely Chern numbers for these two types of systems and show that a light irradiation there can cause Floquet Hamiltonians that contain nodal planes, Weyl points, {multi-Weyl points} or sometimes a completely gapped out spectrum.

\section{Topology}
Let us now discuss the topological characteristics of the nodal surfaces concerned.
For obtaining the corresponding topological charge it is customary to consider a manifold $S^c$ enclosing the nodal surface where all bands are gapped\cite{turker}. The topological invariant is defined on this manifold of lower dimensionality (called codimensionality) $d_c=d-d_{NS}-1$\cite{zhaoprl}. As the dimension of NS, $d_{NS}=2$, the enclosing manifold in a 3D system corresponds to the codimension, $d_c=0$. There are different topological invariants one can look into. One can calculate Chern number in surfaces above and below the NS and take a difference to find the topological charge and consequently the topological stability of the NS\cite{turker}. Again one can consider two points on two sides of the nodal surface and a difference in the estimate of a symmetry preserving perturbation in an eigenstate calculated at two points on two sides of the NS can constitute a topological invariant of the nodal surface\cite{turker}.
We know that the total topological charge should add up to zero within the BZ, as per the no-go theorem\cite{nogo}. So unless there are more than one NSs within the BZ, one can't expect nonzero topological charge for the same.

The Chern number or the Berry phase calculation involves obtaining the closed line integral of Berry connection {about gapless point/points of the spectrum or the corresponding surface integral (as per the Stoke's theorem) of the Berry curvatures\cite{skrev}. For a 3D BZ containing a} NS, we need to consider the 2D cuts above and below the NS and calculate these line and surface integrals there.
The topological charge or Chern numbers of Bloch wavefunctions on a surface enclosing the NS can be obtained using Wilson loop method\cite{yang,chang}.
The Berry connection is given as $\rm A=<\psi_k|-i\nabla_k|\psi_k>$, $|\psi_k>$ being the Bloch wavevectors.
For a chiral symmetric Hamiltonian (\ref{eq1}) with {$c_1=1,~d_1=0$}, one gets $|\psi_k>_\pm=(\frac{\pm\sqrt{4+q_z^2}}{q_z-2i},1)^T$ giving ${\rm A}(q_z)={\rm A}(-q_z)$ leading to zero topological charge for the NS. This is true even considering the full lattice model as well.

For the Hamiltonian (\ref{eq2}) with {$c_2=d_2=1$}, the energy and wavevectors come out to be $E_\pm=\pm sgn(q_z)q_z\sqrt{1+q_\perp^2}$ and $\psi_\pm=[e^{-i\theta}\frac{1\pm sgn(q_z)\sqrt{1+q_\perp^2}}{q_\perp},1]^T$ respectively with $\theta=tan^{-1}(q_y/q_x)$ and $q_\perp^2=q_x^2+q_y^2$. For each band we get $v_z=-v_{-z}$ for $q_z\ne0$. Notice that $\psi_\pm$ and hence $\rm A$ do not depend on $z$. $\rm A$ doesn't depend of $\theta$ either. We get ${\rm A}_z=0$ and $\rm A_\theta=\frac{(1\pm sgn(q_z)\sqrt{1+q_\perp^2})^2}{q_\perp^3}$. The Berry curvature $\Omega=\nabla\times \rm A$ is independent of the magnitude of $q_z$ but contains only the $z$ component:\\ $\Omega=\frac{\mp2sgn(q_z)}{\sqrt{1+q_\perp^2}(1\mp sgn(q_z)\sqrt{1+q_\perp^2})^2}\hat z$.
Interestingly, $\Omega$ blows up at $q_\perp=0$ in the $E_+~(E_-)$ band for $q_z>0~(q_z<0)$ irrespective of the magnitude of $q_z$ (unlike reported in Ref.\cite{xiao}). Considering all possible ${\bf k_0}$ points, one finds singular $\Omega$ in the whole nodal surface as well as above (below) it. Thus it represents a nontrivial topology.

As mentioned before, the calculation of Berry phase involves considering a manifold/surface enclosing the gapless regime of nodal surface. As these nodal surfaces are not simple closed surfaces with genus 0 (like a sphere), we need to consider two surfaces above and below to enclose the nodal surface and then calculate the topological invariant in both of them. Their difference gives the Chern number or Berry phase for the nodal surface.
Due to the symmetry of the problem we consider a cylindrical unit cell and its 2D projection transverse to the $k_z$ direction is a circle in the $q_x-q_y$ plane with radius $2\sqrt{\pi}$. So the Berry flux, which is the line integral over the 2D projection of BZ is given by
\begin{equation}
  \gamma_B=\left(\int q_\perp \rm A_\theta d\theta\right)_{q_\perp=2\sqrt{\pi}}=\frac{(1\pm sgn(q_z)\sqrt{1+4\pi})^2}{2}
  \label{berry-a}
\end{equation}
Notice that it yields $\gamma_B=2\pi$ (and hence Chern number 1) for $q_\perp\rightarrow\infty$, as also mentioned in Ref.\cite{xiao}. But for the present BZ, $\gamma_B|_{q_z\rightarrow 0^+}-\gamma_B|_{q_z\rightarrow 0^-}=\pm2\sqrt{1+4\pi}$ indicate a nonzero Chern number for $E_\pm$ bands signifying existence of topological charge in the nodal surface. {For clarity, we henceforth indicate the Berry fluxes as $\gamma_{B\pm}$ for the $E_\pm$ bands.}
As long as $\Omega$ is nonsingular, one can also calculate the surface integral in the 2D projections as
\begin{eqnarray}
  &&\gamma_{B-}|_{q_z\rightarrow 0^+}=\gamma_{B+}|_{q_z\rightarrow 0^-}=\int_{BZ}\Omega dS=\nonumber\\&-&4\pi\int_0^{2\sqrt{\pi}}\frac{q_\perp dq_\perp}{\sqrt{1+q_\perp^2}(1+\sqrt{1+q_\perp^2})^2}=\frac{1}{2}(1-\sqrt{1+4\pi})^2\nonumber
  \label{berry-b}
\end{eqnarray}
in compatible with Eq.\ref{berry-a}. But $\gamma_{B-}|_{q_z\rightarrow 0^-}$ and $\gamma_{B+}|_{q_z\rightarrow 0^+}$ are not calculable in a similar manner as $\Omega$ blows up at $q_\perp=0$ in those limit. Notice that similar calculations using a straight forward lattice model version of Eq.(\ref{eq1}), namely,
\begin{eqnarray}
H_{L}=\sin q_z(\sin q_x\sigma_x+\sin q_y\sigma_y) +\sin q_z\sigma_z
\label{lat-mod}
\end{eqnarray}
would give a zero Berry phase {which is expected as per the no-go theorem. It is only by inserting topological entities like a Weyl point pair there} ($e.g.,$ in the tight binding model of Eq.\ref{eqtb} as in Ref. \cite{xiao}) one is supposed to get nonzero Chern number although the calculations can be very cumbersome.

{In tune with the bulk-boundary correspondence, a nonzero Chern number in the periodic system indicates that a finite sized NSSM should show topological surface states. We know that a Weyl semimetal (WSM) exhibits Fermi-arc like surface states that connect projections of Weyl points at the boundary\cite{skrev}. Here for a continuum model Hamiltonian $H_{NS}^{(2)}$, one can't expect such Fermi arcs(see Appendix \ref{apA}) though a NSSM containing one NS and a pair of WP within its BZ (as in Hamiltonian (\ref{eqtb})), a finite geometry with boundaries along a transverse direction, say $\hat y$, produces paired surface states that connect, as Fermi-arcs, between different WP and the NS\cite{xiao,yang}.}

\section{Irradiation - Floquet Theory}

The dynamics of a time periodic Hamiltonian is well described using a Floquet mechanism where the time dependent states are expressed via quasi-energies and periodic Floquet modes and the stroboscopic time evolution takes the form of a stationary Hamiltonian called Floquet Hamiltonian\cite{eckardt}. Be it a graphene sheet, a topological insulator or a topological semimetals with time periodic couplings, such Floquet
Hamiltonian often displays nontrivialities different from its time independent counterpart\cite{moessner,yan,debu,banasri}.
{Even Floquet second-order topological insulators can be constructed from a nonsymmorphic crystalline symmetry which is unique in time-periodic systems\cite{peng}.
One can thus very justifiably search for exotic non-equilibrium states in an irradiated NS system} where the time periodic E-M waves couple with the system parameters.

We consider both the cases of Hamiltonian (\ref{eq1}) (with {$c_1=d_1=1$}) and Hamiltonian (\ref{eq2}) to investigate the Floquet theory.
{We know that in an electric field $E$ and the corresponding vector potential $\mathcal{A}$, the modification in the Hamiltonian comes from the Peierl's substitution: $\hbar k\rightarrow\hbar k+e\mathcal{A}$\cite{debu}.}

\subsection{$H_{NS}^{(1)}$ under Irradiation}
  {A time dependent $E$} along $\hat x$ or $\hat y$ does not add any time dependence in the Hamiltonian (\ref{eq1}).
However irradiation via linearly polarized wave of frequency $\omega$ and electric field $E=-E_0cos(\omega t)\hat z$ brings in an vector potential, which in the Coulomb gauge, can be written as $\mathcal{A}=\frac{E_0}{\omega}sin(\omega t)\hat z$. {So for a continuum Hamiltonian (\ref{eq1}), the irradiation results in a time dependent Hamiltonian $H(t)$ given by}
\begin{align}
H(t)&=H^{(1)}_{NS}+\frac{eE_0}{\hbar\omega}sin(\omega t)[q_z\sigma_x-\sigma_y+\sigma_z]\nonumber\\&~~~~~~~~~~~~~~+\frac{e^2E_0^2}{2\hbar^2\omega^2}sin^2(\omega t)\sigma_x
\end{align}
One can see that such a time periodic system leads to an effective Floquet Hamiltonian where no band crossing can appear.
Consider the Fourier modes:
\begin{equation}
  H^{(n)}=\frac{1}{T}\int_0^Tdte^{-in\omega t}H(t)
\end{equation}
$T$ being the time period and $\omega$ the angular frequency of the irradiation.
It leads to $H^{(0)}=H_{NS}^{(1)}+\frac{e^2E_0^2}{4\hbar^2\omega^2}\sigma_x,~~H^{(1)}(H^{(-1)})=-(+)i\frac{eE_0}{2\hbar\omega}[q_z\sigma_x-\sigma_y+\sigma_z]$  and $H^{(n)},H^{-(n)}=0$ for $|n|>1$.
Thus one gets $[H^{(0)},H^{(1)}]=\frac{eE_0}{2\hbar\omega}(q_z^2-\frac{e^2E_0^2}{2\hbar^2\omega^2})[\sigma_y+\sigma_z]$, $[H^{(-1)},[H^{(0)},H^{(1)}]]=\frac{e^2E_0^2}{2\hbar^2\omega^2}(q_z^2-\frac{e^2E_0^2}{2\hbar^2\omega^2})[2\sigma_x+q_z(\sigma_y-\sigma_z)].$\\The effective stationary Hamiltonian at high frequency limit is given by\cite{eckardt}
\vskip -.2 in
\begin{align}
  H_F&=H^{(0)}+\sum_{n>0}\frac{[H^{(n)},H^{-(n)}]}{n\hbar\omega}+\sum_{n\ne0}\frac{[H^{(-n)},[H^{(0)},H^{(n)}]]}{2(n\hbar\omega)^2}\nonumber\\&+\sum_{m\ne n}\frac{[H^{(-n)},[H^{(n-m)},H^{(m)}]]}{3nm(\hbar\omega)^2}+O(\frac{1}{\omega^3})\simeq
  H_{NS}^{(1)}+\nonumber\\&\frac{e^2E_0^2}{4\hbar^2\omega^2}\sigma_x+(\frac{eE_0}{2\hbar^2\omega^2})^2(q_z^2-\frac{e^2E_0^2}{2\hbar^2\omega^2})[2\sigma_x+q_z(\sigma_y-\sigma_z)].
\end{align}
Thus the Floquet system does not see band crossings anymore for the zero of the spectrum would require coefficients of all $\sigma_i$'s to be zero, $i.e.,$ $\frac{q_z^2}{2}+\frac{e^2E_0^2}{4\hbar^2\omega^2}+\frac{e^2E_0^2}{2\hbar^4\omega^4}(q_z^2-\frac{e^2E_0^2}{2\hbar^2\omega^2})=0$ and $q_z[1-\frac{e^2E_0^2}{8\hbar^4\omega^4}(2q_z^2-\frac{e^2E_0^2}{\hbar^2\omega^2})]=0$ {simultaneously. But no real $q_z$ can satisfy such criteria.} The Floquet system thus features a gapped spectrum {with no topological surface states in a system with finite geometry.}

\subsection{$H_{NS}^{(2)}$ under Irradiation}

As we did not consider any $k_x$ or $k_y$ dependence in the Hamiltonian (\ref{eq1}), the variation in this Floquet system is not rich enough. But when we consider Hamiltonian (\ref{eq2}), interesting dynamic behavior can be witnessed.

\subsubsection{Linear Polarization}
In this case, for a linearly polarized light with $E=-E_0cos(\omega t)\hat x$ or $\mathcal{A}=\frac{E_0}{\omega}sin(\omega t)\hat x$ the time dependent Hamiltonian becomes
\begin{eqnarray}
H(t)=H_{NS}^{(2)}+\frac{eE_0}{\hbar\omega}sin(\omega t){c_2}q_z\sigma_x .
\end{eqnarray}
Among the time-Fourier modes we get $H^{(0)}=H_{NS}^{(2)},~~H^{(1)}(H^{(-1)})=-(+)i\frac{eE_0}{2\hbar\omega}{c_2}q_z\sigma_x$  and $H^{(n)},H^{-(n)}=0$ for all integer $|n|>1$. So the 1st order correction goes to zero.
Then one gets $[H^{(0)},H^{(1)}]={c_2}q_z^2\frac{eE_0}{\hbar\omega}[{d_2}\sigma_y-{c_2}q_y\sigma_z]$ and consequently $[H^{(-1)},[H^{(0)},H^{(1)}]]=-{c_2}^2q_z^3(\frac{eE_0}{\hbar\omega})^2[{d_2}\sigma_z+{c_2}q_y\sigma_y]$. So up to the 2nd order, the Floquet Hamiltonian thus takes the form\cite{eckardt}
\begin{align}
  H_F\simeq  H_{NS}^{(2)}-{c_2}^2q_z^3\frac{(eE_0)^2}{2(\hbar\omega)^4}[{d_2}\sigma_z+{c_2}q_y\sigma_y]
  \label{eqhf}
\end{align}
which indicates a pair of additional {band-crossing} points at {$q=(0,0,\pm\sqrt{2}\frac{(\hbar\omega)^2}{{c_2}eE_0}$) or $k=(k_{x0},k_{y0},\mp\pi\pm\sqrt{2}\frac{(\hbar\omega)^2}{{c_2}eE_0}$). They constitute multi-Weyl points\cite{multiwsm}} (in case they fall within the BZ boundary despite the high frequency limit, {$i.e.,$ for $\sqrt{2}\frac{(\hbar\omega)^2}{{c_2}eE_0}<\pi$) as the dispersion along $q_x$ and $q_y$ directions become nonlinear there. At this point we should pause to think back that our rationale to use the continuum model $H_{NS}^{(2)}$ is to study the system within a small energy window about the NS energy at the point $k_0=(\frac{4\pi}{3},0,\pi)$. If we would decide to study about a different point, say $k_0=(0,0,\pi)$ (see Appendix \ref{apB}), we would not get this multi-WP pair in the Floquet spectrum.}

{Going back again to the Floquet Hamiltonian (\ref{eqhf}), we conjecture that for such interesting outcomes of having 1 NS and 2 WPs and 2 multi-WP within the Floquet BZ, quite nontrivial surface states can be expected. We need to however resort to full wave simulations\cite{xiao} using COMSOL package\cite{comsol} to probe it further numerically.}

Similarly, if we change the angle of polarization resulting in $\mathcal{A}=\frac{E_0}{\omega}sin(\omega t)\hat z$, we get
\begin{eqnarray}
H(t)=H_{NS}^{(2)}+\frac{eE_0}{\hbar\omega}sin(\omega t)[{c_2}(q_x\sigma_x+q_y\sigma_y)+{d_2}\sigma_z]\nonumber\\
\label{eq}
\end{eqnarray}
In this case the energy zeros corresponding to $H(t)$ appear at $(qx,qy,-\frac{eE_0}{\hbar\omega}sin(\omega t))$, showing the nodal plane to sinusoidally fluctuate in time with $q_z$ varying between ($-\frac{eE_0}{\hbar\omega}:\frac{eE_0}{\hbar\omega}$). But the high frequency expansion does not lead to any new terms and the Floquet Hamiltonian remains at the bare zeroth level as: $H_F=H_{NS}^{(2)}$. 

\subsubsection{Circular Polarization}
Next we consider a circularly polarized light for which we take $\mathcal{A}=\frac{E_0}{\omega}[sin(\omega t)\hat x+cos(\omega t)\hat y]$ and one gets 
\begin{eqnarray}
H(t)&=&H_{NS}^{(2)}+\frac{eE_0}{\hbar\omega}q_z{c_2}[sin(\omega t)\sigma_x+cos(\omega t)\sigma_y].\nonumber\\
\label{eq2c}
\end{eqnarray}
Again we obtain $H^{(0)}=H_{NS}^{(2)}$. But now $H^{(1)}(H^{(-1)})=\frac{eE_0}{2\hbar\omega}q_z{c_2}[-(+)i\sigma_x+\sigma_y]$. Thus $[H^{(1)},H^{(-1)}]=(\frac{eE_0}{\hbar\omega})^2q_z^2{c_2}^2\sigma_z$. And we obtain nonzero correction even in the 1st order of the expansion.
More precisely, the Floquet Hamiltonian is given by,
\begin{eqnarray}
  H_F&=&H^{(0)}+\sum_{n>0}\frac{[H^{(n)},H^{-(n)}]}{n\hbar\omega}+O(\frac{1}{\omega^2})\nonumber\\
  &\simeq&H_{NS}^{(2)}+\frac{(eE_0q_z{c_2})^2}{\hbar^3\omega^3}\sigma_z
\end{eqnarray}
\\
Again this time, the Floquet Hamiltonian indicates an additional single multi-WP at $q=(0,0,-\frac{{d_2}\hbar^3\omega^3}{(eE_0{c_2})^2})$ in the high frequency limit (but this ceases to exist for a different choice of $k_0=(0,0,\pi)$ within the NS). {Thus the BZ now becomes an assembly of 1 NS and 2 WPs and 1 multi-WP whose Chern numbers should be compatible with the no-go theorem that ensures absence of overall topological charges in the system. Also one can expect interesting topological surface states if there are finite boundaries in the system.}

More generally, for an elliptically polarized light with $\mathcal{A}=\frac{E_0}{\omega}[sin(\omega t+\phi)\hat x+cos(\omega t)\hat y]$ we get
\begin{eqnarray}
H(t)&=&H_{NS}+\frac{eE_0}{\hbar\omega}q_z{c_2}[sin(\omega t+\phi)\sigma_x+cos(\omega t)\sigma_y]\nonumber\\
\label{eq2c}
\end{eqnarray}
Again $H^{(0)}=H_{NS}^{(2)}$. But now $H^{(1)}(H^{(-1)})=\frac{eE_0}{2\hbar\omega}q_z{c_2}[-(+)ie^{+(-)i\phi}\sigma_x+\sigma_y]$. Thus $[H^{(1)},H^{(-1)}]=(\frac{eE_0}{\hbar\omega})^2q_z^2{c_2}^2cos(\phi)\sigma_z$.
So other than the nodal plane at $q_z=0$ we get a {multi-Weyl point} at $q_z=-\frac{{d_2}\hbar^3\omega^3}{(eE_0{c_2})^2cos\phi}$, if that falls within the BZ.
Thus this {additional Weyl point} recedes more from the BZ boundary at $k_z=\pi$ as $\phi$ is gradually increased, until it crosses the opposite boundary of the BZ for $\cos\phi=\frac{\pi(eE_0{c_2})^2}{{d_2}\hbar^3\omega^3}$.

\subsection{{Irradiation on a dispersive NS}}

{The Hamiltonian $H_{NS}^{(2)}$ is a generalized version of a continuum model derived from the tight-binding Hamiltonian (\ref{eqtb}) (as shown in Ref.\cite{xiao}) when expanded about the point $k_0=(\frac{4\pi}{3},0,\pi)$. But we must not omit the dispersions within the NS in these models. One can also consider an expansion about $k_0=(0,0,\pi)$ where the dispersion shows a minima (see Fig.\ref{disp}). In Appendix \ref{apB}, we have shown how energy dependent nodal surfaces can be obtained from tight-binding Hamiltonian (\ref{eqtb}) for $k_0=(\frac{4\pi}{3},0,\pi)$ as well as $k_0=(0,0,\pi)$.
Similar to $H_{NS}^{(2)}$, one such generalized Hamiltonian can be written as
\begin{eqnarray}
H_{NS}^{(3)}=[c_3q_\perp^2+d_3q_z^2]{I}_2+H_{NS}^{(2)},
\label{eqh3}
\end{eqnarray}
where the NS is given by $q_z=0$, characterized by a $q_\perp$ dependent energy profile.}

{The effect of irradiation on such $q$-dependent diagonal term is worth inspecting. Notice that the additional diagonal term only contributes to the zeroth-order term in the Floquet Hamiltonian, obtained from the high frequency expansion. For linear polarization having $\mathcal{A}=\frac{E_0}{\omega}sin(\omega t)\hat x$, the Peierl's substitution causes such zeroth order term to be
  \begin{equation}
    H^{(0)}=H_{NS}^{(3)}+\frac{c_3}{2}(\frac{eE_0}{\hbar\omega})^2{I}_2.
    \end{equation}
We have already seen that for similar irradiation on a Hamiltonian $H_{NS}^{(2)}$, the Floquet spectrum exhibits a multi-WP pair and one NS at $q_z=0$. In the present case, it only modifies the diagonal term and accordingly we obtain:
  \begin{align}
    H_F\simeq  H_{NS}^{(3)}+\frac{c_3}{2}(\frac{eE_0}{\hbar\omega})^2{I}_2-{c_2}^2q_z^3\frac{(eE_0)^2}{2(\hbar\omega)^4}[{d_2}\sigma_z+{c_2}q_y\sigma_y]\nonumber\\
\label{eqe3}
  \end{align}}
  
  \begin{figure}
    \vskip .1 in
  \begin{center}
   \begin{picture}(100,100)
     \put(-70,0){
    \includegraphics[width=.5\linewidth,height=1.4 in]{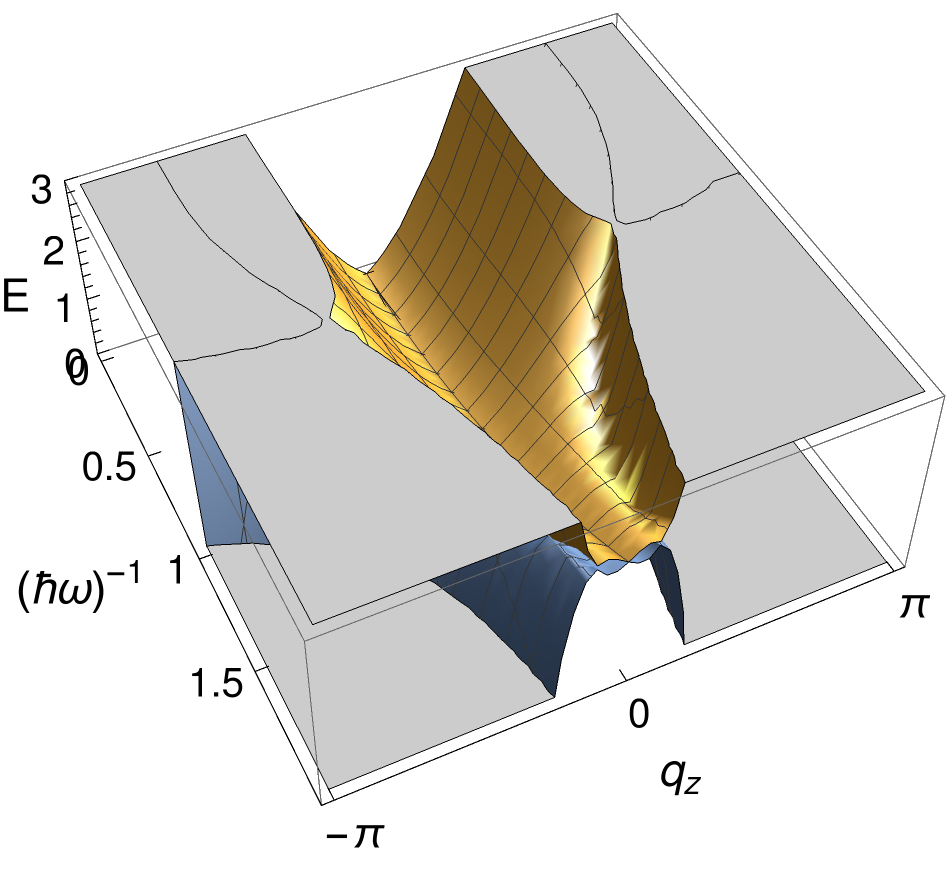}\hskip .1 in
    \includegraphics[width=.45\linewidth,height= 1.4 in]{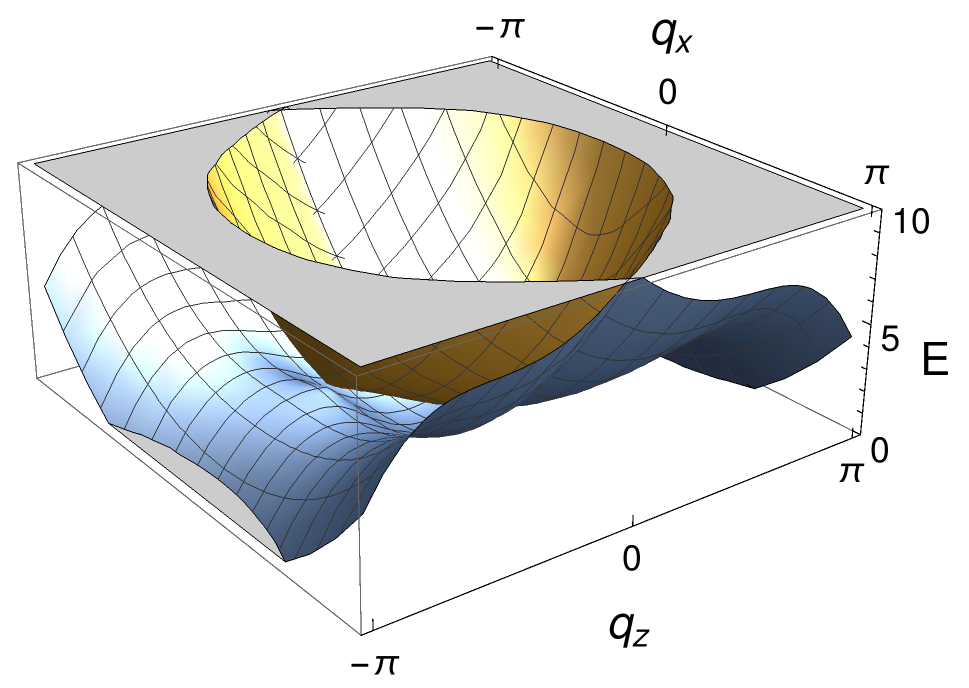}
    \put(-250,90){(a)}
     \put(-120,90){(b)}}
   \end{picture}
   \end{center}
   \vskip -0.2 in
\caption{{Low energy dispersions from the Floquet model (\ref{eqe3}) for (a) $q_x=q_y=0$ and (b) $q_y=0,~\hbar\omega=1$. Nodal surfaces are obtained at $q_z=0$ and $q_z=\pm\sqrt{2}\frac{(\hbar\omega)^2}{c_2eE_0}$. Here we consider $c_2=d_2=c_3=d_3=1$ and in the units of $eE_0=1$.}} 
   \label{disp2}
\end{figure}
{Similar results can be obtained for polarization along $\hat z$ direction having $\mathcal{A}=\frac{E_0}{\omega}sin(\omega t)\hat z$.}


{In Fig.\ref{disp2}, we show the spectral variations with $q_z$ corresponding to the Hamiltonian (\ref{eqe3}) for $q_y=0$ and different values for $q_x$ and $(\hbar\omega)^{-1}$ (these dispersions, however, are valid in the small $q$ limit and at low energies about the NS at $q_z=0$). Notice the dispersive nature of the nodal surface at $q_z=0$. Apart from this NS at $q_z=0$ the Floquet spectrum also shows a pair of multi-WPs (corresponding to each $\omega$ shown there) within the BZ\cite{cmnt} where the two bands touch.}

\section{Summary}
In this present work, we have studied simple two band models of NSSM to understand its low energy behavior near the band crossings. While some nodal surfaces appear due to additional internal symmetries and carry $\mathbb{Z}_2$ topological index with them, some others, not necessarily topologically charged, are outcome of composite symmetries including nonsymmorphic symmetry that make the nodal surface essentially pinned at the BZ sub-boundaries. Charged NS, in compatible with the no-go theorem can be designed in acoustic metamaterials\cite{xiao} with the presence of additional Weyl points within the 3D BZ. Under irradiation, such systems, both with or without a topological charge, show interesting time periodic behavior which for stroboscopic time variation can be well analyzed using Floquet-Magnus theory. {We find that the nodal surface ceases to exist in the Floquet spectrum that arises from a NSSM Hamiltonian containing NS of zero topological charge.} But in presence of a charged NS under different choice of irradiation polarizations, {one can witness a plethora of interesting dynamical responses that can add no additional feature in the Floquet spectrum or can introduce additional multi Weyl points in the Floquet spectrum demanding exotic surface state features in a system with finite boundaries. Notice that the NSSM systems or their Floquet versions have topological protections as long as NS or WPs are present in them. But had we considered open quantum systems, dissipative effects on the Floquet models would also have been important that have the potential to break such topological robustness\cite{oqs}.Though not considered in this paper, we plan to venture on that path in a future communication.}

Nodal surfaces are not necessarily equi-energy surfaces but are of more practical interest when situated close to Fermi levels. {We also consider dispersive nodal surfaces where energy varies within the surface itself. By merely tuning the Hamiltonian parameters, one can flatten or make the NS more dispersive.} Our calculations offer scopes to delve into further deep analysis on dynamic behavior of synthesized materials exhibiting nodal planes close to Fermi surfaces.  For example, de Haas-van Alphen spectroscopy in combination with density function theory calculations can show nonsymmorphic symmetry in chiral, ferromagnetic compound $MnSi$ and consequent nodal planes enforces topological protectorates with large $\Omega$ at the intersection of NS and the Fermi surface\cite{wilde}. An irradiation on such system and a consequent Floquet analysis can be performed to explore what dynamic features it has to offer. {One can digitally witness the Floquet phases using programmable superconducting quantum processors\cite{digit}.} Furthermore, it will be interesting to probe the effect of time periodic perturbation on the semimetal $ZrSiS$ where a coexistence of nodal planes and nodal lines are observed\cite{zrsis} at low energies. Floquet nodal rings and Floquet nodal spheres have also been studied using four-band theories where topological invariants are captured using quantum metric measurements\cite{metric}.
Similar calculations can be repeated for the non-symmorphic symmetry protected nodal surfaces pinned at Floquet BZ sub-boundaries. Such time periodic drives have the potential to design tunable topological qubits\cite{yu}.

{As nodal surfaces are getting more and more popular in the condensed matter community, with still not very comprehensive analysis out yet of its models or effect subjected to irradiation, our work attempts to provide useful feedback in that pool of search. As for the scope of improvement, we admit that the outcomes presented/discussed in this paper are based on continuum model calculations about the NS and are not meant to truly describe the physics in the whole BZ. This also hinders us to comment more exclusively on the surface states. In future, we plan to perform a full scale calculation using the tight-binding Hamiltonian and simulate the same using COMSOL\cite{comsol} to understand better the surface states in a system with finite geometry or to its Floquet version.}
\\

\section*{Acknowledgement}
SK {thanks D. Sinha for fruitful discussions and} acknowledges financial support from DST-SERB, Government of India via grant no. CRG/2022/002781.

\appendix
\section{{Surfaces states corresponding to $H_{NS}^{(2)}$}}\label{apA}
{Consider a finite sized NSSM model represented by Eq.(\ref{eq2}) for $|y|<y_0$ with boundaries at $|y|=y_0$. Following the approach shown in Okugawa $et.~al.$\cite{murakami} with $q_y$ replaced by $-i\partial_y$ in the Hamiltonian, the bound states at the boundaries can be given by  $E_\pm=\pm sgn(q_z) c_2q_zq_x$ and $\psi_\pm=[\frac{d_2\pm sgn(q_z)c_2 q_x}{c_2q_x-d_2},1]^T\Theta(y_0-|y|)e^{-\frac{d_2}{c_2} (y\mp y_0)}$. Notice that though it indicates zero energy surface states for $q_z=0$ or $q_x=0$, one can only expect Fermi arcs (for $\mu=0$) had we considered full tight binding model including the WP pair as well\cite{xiao}. }

\section{{Dispersions within a nodal surface}}\label{apB}

{Let us derive the continuum model coming out of the tight-binding Hamiltonian (\ref{eqtb}). Considering $k_0=(4\pi/3,0,\pi)$ and writing $k=k_0+q$, one obtains
$h_1(\pi+q_z)=\tilde h_1(q_z)=2t_c[\cos( \frac{4\pi}{3} + q_x+\pi+q_z)+2\cos(\frac{2\pi}{3}+\frac{q_x}{2}-\pi-q_z) \cos(\frac{\sqrt{3}}{2}q_y)]$. So for small $q$ this becomes
\begin{align}
\tilde h_1(q_z)&=2t_c[ \cos( \frac{7\pi}{3}) \cos( q_x+q_z) - \sin( \frac{7\pi}{3}) \sin( q_x+q_z) +\nonumber\\
2&\{\cos \frac{\pi}{3} \cos( \frac{q_x}{2} - q_z) + \sin\frac{\pi}{3} \sin( \frac{q_x}{2} - q_z)\}\cos\frac{\sqrt{3}q_y}{2}]\nonumber\\
=&2t_c[ \frac{1}{2} ( 1-\frac{(q_x+q_z)^2}{2}) -\frac{\sqrt{3}}{2} (q_x+q_z)+ \{  1-\nonumber \\
& \frac{(\frac{q_x}{2}-q_z)^2}{2}+ \sqrt{3} ( \frac{q_x}{2}-q_z)\}( 1-\frac{3q_y^2}{8})]\nonumber\\
&= 3t_c - \frac{3}{4} ( q_\perp^2+ 2q_z^2) -3 \sqrt{3}t_c q_z+O(q^4).
\end{align}}

{Similarly the off-diagonal term modifies to,
\begin{align}
  h_2&=2t_0 \cos( \frac{\pi}{2}+\frac{q_z}{2})[ 2\cos( \frac{2\pi}{3}+\frac{q_x}{2})\{ \cos(\frac{\sqrt{3}q_y}{6})\nonumber\\
    &+ i \sin(\frac{\sqrt{3}q_y}{6})\} + \cos(\frac{\sqrt{3}q_y}{3}) - i \sin(\frac{\sqrt{3}q_y}{3})] \nonumber\\
&=-2t_0( \frac{q_z}{2})[ \{ 2 ( -\frac{1}{2}) ( 1-\frac{q_x^2}{8}) - \frac{\sqrt{3}q_x}{2} \} ( 1-\frac{q_y^2}{24}  \nonumber\\& + i \frac{\sqrt{3}q_y}{6}) +( 1-\frac{q_y^2}{6} - i \frac{\sqrt{3}q_y}{3}) ]\nonumber\\
  &=\frac{\sqrt{3}}{2}t_0q_z ( q_x  + i q_y )+O(q^3)\\
  &\rm{So~the~continuum~Hamiltonian~becomes}\nonumber\\
H&= 3t_c[ 1 - \frac{1}{4} ( q_\perp^2 + 2q_z^2)]I +\frac{\sqrt{3}}{2}[ t_0q_z ( q_x \sigma_x - i q_y \sigma_y)\nonumber\\&\mp 6t_cq_z\sigma_z]+O(q^3).
\end{align}}

{On the other hand, writing $k=k_0+q$ with $k_0=(0,0,\pi)$ in Eq.(\ref{eqtb}), one gets 
$\tilde h_1(q_z)=2t_c[\cos( q_x+\pi+q_z)+2\cos(\frac{q_x}{2}-\pi-q_z) \cos(\frac{\sqrt{3}}{2}q_y)]$.}

{So for small $q$, we get
\begin{align}
  \tilde h_1(q_z)&=2t_c[ \cos\pi \cos( q_x+q_z) - \sin\pi \sin( q_x+q_z) +\nonumber\\
    &2 \{ \cos \pi \cos( \frac{q_x}{2}-q_z) + \sin \pi  \sin( \frac{q_x}{2}-q_z)\} \cos( \frac{\sqrt{3}}{2}q_y)]\nonumber\\
  =&2t_c[-( 1-\frac{(q_x+q_z)^2}{2})-2( 1-\frac{(q_x/2-q_z)^2}{2}) ( 1-\frac{3q_y^2}{8})]\nonumber\\
&=2t_c[ -3 + \frac{3}{4} q_x^2+\frac{3}{4} q_y^2+\frac{3}{2} q_z^2]+O(q^4).
\end{align}}

{Similarly the off-diagonal term modifies to,
\begin{align}
  h_2&=2t_0 \cos( \frac{\pi}{2}+\frac{q_z}{2})[ 2\cos( \frac{q_x}{2})\{ \cos(\frac{\sqrt{3}q_y}{6})\nonumber\\
    &+ i \sin(\frac{\sqrt{3}q_y}{6})\} + \cos(\frac{\sqrt{3}q_y}{3}) - i \sin(\frac{\sqrt{3}q_y}{3})] \nonumber\\
&= -2t_0( \frac{q_z}{2})[ 2( 1-\frac{q_x^2}{8})( 1-\frac{q_y^2}{24}   + i \frac{\sqrt{3}q_y}{6}) \nonumber\\&+( 1-\frac{q_y^2}{6} - i \frac{\sqrt{3}q_y}{3}) ]\nonumber\\
&= -3t_0q_z+O(q^3).
\end{align}
Hence the continuum Hamiltonian becomes
\begin{equation}
H=-6t_c+3(\frac{q_\perp^2}{2}+q_z^2){I}_2-3t_0q_z\sigma_x +O(q^3).
  \end{equation}}


\begin{thebibliography}{99}
 
\bibitem{wu} W. Wu $et~al.$, Phys. Rev. B{\bf 97}, 115125 (2018).
\bibitem{zhong} C. Zhong $et~al.$, Nanoscale {\bf 8}, 7232 (2016).
     \bibitem{turker} O. Turker $et~al.$, Phys. Rev. B{\bf 97}, 075120 (2018).
     \bibitem{1807} P. A. volkov, and S. Moroz, Phys. Rev. B{\bf 98}, 241107 (2018).
     \bibitem{wang} J. Wang $et~al.$, Phys. Rev. B{\bf 98}, 201112(R) (2018).
     \bibitem{liang} Q.-F. Liang $et~al.$, Phys. Rev. B {\bf 93}, 085427 (2016).
     \bibitem{timm} Q.-F. Liang $et~al.$, Phys. Rev. Lett. {\bf 118}, 127001 (2017).
  \bibitem{xiao} M. Xiao $et~al.$, arXiv:1709.02363 (2017); M. Xiao $et~al.$, Sci. Adv.{\bf 6}, eaav2360 (2020).
   \bibitem{yang} Y. Yang $et~al.$, Nat. com. {\bf 10}, 5185 (2019).
  \bibitem{furusaki} A.Furusaki, Science Bulletin {\bf 62}, 788-794 (2017).
   \bibitem{zhao} Y. X. Zhao $et~al.$, Phys. Rev. B{\bf 94}, 195109 (2016).
   \bibitem{nogo} H. Nielsen, and M. Ninomiya, Nuc. Phys. B {\bf 193}, 173 (1981).
   \bibitem{eckardt} A. Eckardt $et~al.$, New Jour. Phys. {\bf 17}, 093039 (2015).
     \bibitem{zhaoprl} Y. X. Zhao $et~al.$, Phys. Rev. Lett.{\bf 116} 156402 (2016).
\bibitem{skrev} {S. Kar, A. Jayannavar, Asian Jour. of Res. and Rev. in Phys., {\bf 4(1)}, 34-45 (2021).}
   \bibitem{chang} G. Chang $et~al.$, Nat. Mat. {\bf 17}, 978 (2018).
    \bibitem{moessner} J. Cayssol $et~al.$, Phys. Stat. Solidi RRL{\bf 7}, No.1-2, 101 (2013).
 \bibitem{yan} Z. Yan $et~al.$, Phys. Rev. Lett.{\bf 117}, 087402 (2016).
\bibitem{debu} D. Sinha $et~al.$, Cur. App. Phys.{\bf 18}, 1087 (2018).
\bibitem{banasri} S. Kar $et~al.$, Phys. Rev. B{\bf 98}, 245119 (2018).
\bibitem{peng} {Y. Peng $et~al.$, Phys. Rev. Lett.{\bf 123}, 016806 (2019).}        
   \bibitem{multiwsm} {C. Fang $et~al.$, Phys. Rev. Lett.{\bf 108}, 266802 (2012).}
   \bibitem{comsol} {COMSOL Multiphysics ver. 5.2. www.comsol.com. COMSOL AB.}
   \bibitem{cmnt}  {Notice that the original Weyl points are absent in the spectrum as they are far from $k_0$ and we don't expect this continuum model to produce true dispersions there. In future, we plan to do a more rigorous finite size simulation using a COMSOL software that will more exactly describe both the bulk as well as the boundary spectrum.}
       \bibitem{oqs} {S. A. Sato $et~al.$, Jour. Phys. B: At. Mol. Opt. Phys., {\bf 53}, 225601 (2020).}
       \bibitem{wilde} A. Wilde $et~al.$, Nature {\bf 594}, 374 (2021).
       \bibitem{digit} {X. Zhang $et~al.$, Nature {\bf 607}, 468 (2022).}
  \bibitem{zrsis} B. -B. Fu $et~al.$, Sci. Adv. {\bf 5}, eaau6459 (2019).
 \bibitem{metric} G. Salerno $et~al.$, Phys. Rev. Res.{\bf 2}, 013224 (2020).
 \bibitem{yu} M. Yu $et~al.$, Natl. Sc. Rev., {\bf 7}, 254-260 (2020).
  {\bibitem{murakami} R. Okugawa $et~al.$, Phys. Rev. B{\bf 89}, 235315 (2014).}
  
  

\end{thebibliography}
\end{document}